\documentclass[prd,preprint,showpacs,preprintnumbers,nofootinbib,eqsecnum,superscriptaddress]{revtex4}

 \usepackage[dvips,final]{graphicx}
  \usepackage{amssymb}
   \usepackage{amsmath}
    \usepackage{amsfonts}
     \usepackage{epsfig}
      \usepackage{bm}

\usepackage{mathpazo}

\usepackage[section]{placeins}

\usepackage{multirow}
\usepackage{ctable}
\usepackage{booktabs}
\usepackage{array}
\usepackage{tabularx}
\usepackage{xcolor}
\usepackage{pstricks}

\begin{document}

\title{Production of two charm quark-antiquark pairs\\
in single-parton scattering 
within the \bm{$k_t$}-factorization approach}

\author{Andreas van Hameren}
\email{andreas.hameren@ifj.edu.pl}
\affiliation{Institute of Nuclear Physics PAN, PL-31-342 Cracow, Poland}

\author{Rafa{\l} Maciu{\l}a}
\email{rafal.maciula@ifj.edu.pl} 
\affiliation{Institute of Nuclear Physics PAN, PL-31-342 Cracow, Poland}

\author{Antoni Szczurek}
\email{antoni.szczurek@ifj.edu.pl}
\affiliation{Institute of Nuclear Physics PAN, PL-31-342 Cracow, Poland and\\
University of Rzesz\'ow, PL-35-959 Rzesz\'ow, Poland}

\date{\today}

\begin{abstract}

We present first results for the $2 \to 4$ single-parton scattering $g g \to c \bar c c \bar c$ subprocess for the first time fully within
the $k_t$-factorization approach. In this calculation we have used
the Kimber-Martin-Ryskin unintegrated gluon distribution which effectively
includes some class of higher-order gluon emissions, and an off-shell
matrix element squared calculated using recently developed techniques.
The results are compared with our earlier result obtained within the collinear-factorization approach.
Only slightly larger cross sections are obtained than in the case
of the collinear approach.
Inclusion of transverse momenta of gluons entering the hard process
leads to a much stronger azimuthal decorrelation between $c c$ and
$\bar c \bar c$ than in the collinear-factorization approach.
A comparison to predictions of double parton scattering (DPS) results
and the LHCb data strongly suggests that the assumption of two
fully independent DPS ($g g \to c \bar c \otimes g g \to c \bar c$)
may be too approximate.

\end{abstract}


\maketitle

\section{Introduction}
 
At high energy, gluon-gluon fusion becomes the dominant
mechanism of heavy $c \bar c$ or $b \bar b$ pair production.
The cross section for single pair production can be calculated
either in collinear next-to-leading order approach or
the $k_t$-factorization approach. The $Q \bar Q$ and Higgs production
are golden reactions for applications of the $k_t$-factorization
approach \cite{CCH91,CE91,BE01,Teryaev,Baranov00,Zotov:2003cb,BLZ,Luszczak:2005cq,Shuvaev,LMS09,MSS2011,Pasechnik:2006du,Lipatov:2014mja,Szczurek:2014mwa}.
In the $k_t$-factorization
approach the basic ingredients are so-called unintegrated
gluon distribution functions (UGDFs) and off-shell matrix elements. 
Different models of UGDFs have been proposed in the literature. 
The Kimber-Martin-Ryskin (KMR) \cite{KMR} UGDF is believed to include 
the dominant higher-order corrections. 
The off-shell matrix elements for $g g \to Q \bar Q$ were calculated
already long ago \cite{CCH91,CE91,BE01}.
The $k_{t}$-factorization formalism was applied recently in the context of experimental
data measured at the LHC \cite{JKLZ,Saleev:2012np,Maciula:2013wg,Nefedov:2014qea,Karpishkov:2014epa}
and a relatively good description was obtained when using the KMR UGDF.

In the case of the Higgs boson production both $2 \to 1$ and $2 \to 2$ subprocess
have to be taken into account \cite{Szczurek:2014mwa}. 
In Ref.~\cite{Baranov:2008rt} a $2 \to 3$ $g g \to c \bar c \gamma$ subprocess
was taken into account when calculating cross sections for 
$p p \to c \bar c \gamma X$ reaction.
Recently the $k_t$-factorization approach was also applied to 
three-jet \cite{vanHameren:2013fla} and $Zb\bar{b}$ \cite{Zotov2015} production.

A convenient formalism for the automation of the calculation of tree-level scattering amplitudes with off-shell gluons for arbitrary processes was recently introduced in Ref.~\cite{vanHameren:2012if}.
Off-shell gluons are replaced by eikonal quark-antiquark pairs, and the amplitude can be calculated with the help of standard local Feynman rules, including the eikonal gluon-quark-antiquark vertex and the eikonal quark-antiquark propagator.
The well-known successful recursive methods to calculate tree-level amplitudes can directly be applied, including the ``on-shell'' recursion, or Britto-Cachazo-Feng-Witten recursion, as shown in Ref.~\cite{vanHameren:2014iua}.
The heuristic introduction of the formalism in Ref.~\cite{vanHameren:2012if} has be given solid ground in Ref.~\cite{Kotko:2014aba}.
Most of the afford was devoted to dijet production \cite{vanHameren:2014lna,vanHameren:2014ala} so far.

The $p p \to c \bar{c} c \bar{c} X$ reaction is interesting by itself.
It was shown by us recently that this reaction is a golden reaction
to study double-parton scattering (DPS) processes \cite{Luszczak:2011zp,Maciula:2013kd}.
The LHCb collaboration confirmed the theoretical predictions
and obtained a large cross section for production of two mesons,
both containing $c$ quarks or both containing $\bar c$ antiquarks \cite{Aaij:2012dz}.
The single-parton scattering (SPS) contribution was discussed in Refs.~\cite{Schafer:2012tf} and \cite{vanHameren:2014ava}.
In the first case \cite{Schafer:2012tf} a high-energy approximation 
was used neglecting some unimportant at high energies Feynman diagrams. Last year we have calculated
the lowest-order SPS cross section(s) including a complete set of Feynman 
diagrams \cite{vanHameren:2014ava} in the collinear-factorization approach.
The final result was only slightly different than that obtained
in the high-energy approximation.

In the present letter we wish to go one step further and try to
calculate the SPS cross sections for the $p p \to c \bar c c \bar c X$ reaction 
consistently in the $k_t$-factorization approach. Doing so we may hope 
that a sizeable part of higher-order corrections will be included.
On the technical side this will be a first calculation
within the $k_t$-factorization approach based on a $2 \to 4$ subprocesses
with two off-shell initial-state partons (gluons).
The result is also important in the context of studying DPS as 
the considered SPS mechanism constitutes an irreducible background, and its
estimation is therefore of prior importance if deeper conclusions
concerning DPS can be drawn from measurements at the LHC.

\section{Formalism}

\begin{figure}[!h]
\begin{minipage}{0.35\textwidth}
 \centerline{\includegraphics[width=1.0\textwidth]{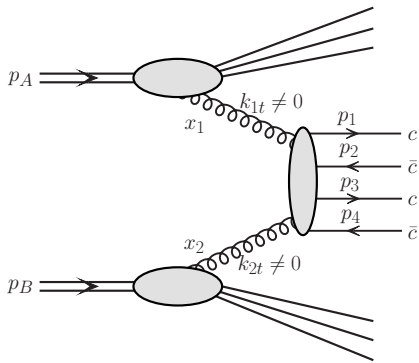}}
\end{minipage}
\caption{
\small A diagrammatic representation of the considered mechanism of $c \bar c c \bar c$ final-state production via single-parton scattering within $k_{t}$-factorization approach.
}
 \label{fig:mechanism}
\end{figure}

Within the $k_t$-factorization approach the SPS cross section for 
$p p \to c \bar c c \bar c X$ reaction, sketched in Fig.~\ref{fig:mechanism}, can be written as
\begin{equation}
d \sigma_{p p \to c \bar c c \bar c} =
\int d x_1 \frac{d^2 k_{1t}}{\pi} d x_2 \frac{d^2 k_{2t}}{\pi}
{\cal F}(x_1,k_{1t}^2,\mu^2) {\cal F}(x_2,k_{2t}^2,\mu^2)
d {\hat \sigma}_{gg \to c \bar c c \bar c}
\; .
\label{cs_formula}
\end{equation}
In the formula above ${\cal F}(x,k_t^2,\mu^2)$ are unintegrated
gluon distributions that depend on longitudinal momentum fraction $x$,
transverse momentum squared $k_t^2$ of the gluons entering the hard process,
and in general also on a (factorization) scale of the hard process $\mu^2$.
The elementary cross section in Eq.~(\ref{cs_formula}) can be written
somewhat formally as:
\begin{eqnarray}
d {\hat \sigma} &=&
\frac{d^3 p_1}{2 E_1 (2 \pi)^3} \frac{d^3 p_2}{2 E_2 (2 \pi)^3}
\frac{d^3 p_3}{2 E_3 (2 \pi)^3} \frac{d^3 p_4}{2 E_4 (2 \pi)^3}
(2 \pi)^4 \delta^{4}(p_1 + p_2 + p_3 + p_4 - k_1 - k_2) \nonumber \\
&&\times\frac{1}{\mathrm{flux}} \overline{|{\cal M}_{g^* g^* \to c \bar c c \bar c}(k_{1},k_{2})|^2}
\; ,
\label{elementary_cs}
\end{eqnarray}
where only dependence of the matrix element on four-vectors of gluons $k_1$ and $k_2$ 
is made explicit. In general all four-momenta associated with partonic legs enter.
The matrix element takes into account that both gluons entering the hard
process are off-shell with virtualities 
$k_1^2 = -k_{1t}^2$ and $k_2^2 = -k_{2t}^2$.
The matrix element squared is rather complicated
and explicit formula will be not given here.

\begin{figure}
\begin{center}
\epsfig{figure=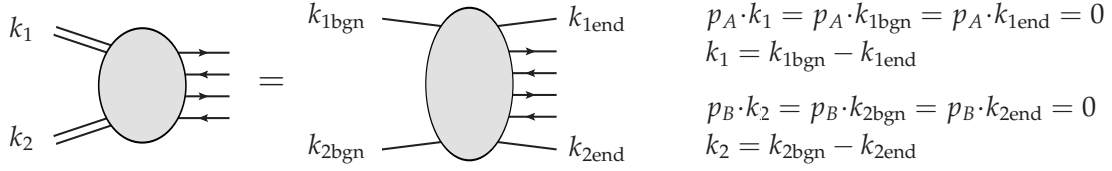,width=0.9\linewidth}
\caption{\label{kinematics} Momenta of the off-shell gluons, represented as double lines on the left hand side, and the eikonal quark-antiquark pairs. The amplitude is independent of a simultaneous momentum shift $k_{1\mathrm{bgn}}+q$, $k_{1\mathrm{end}}+q$ as long as $p_A\!\cdot\!q=0$. The same holds for the other eikonal line with $p_B$.}
\end{center}
\end{figure}
\begin{figure}
\begin{center}
\epsfig{figure=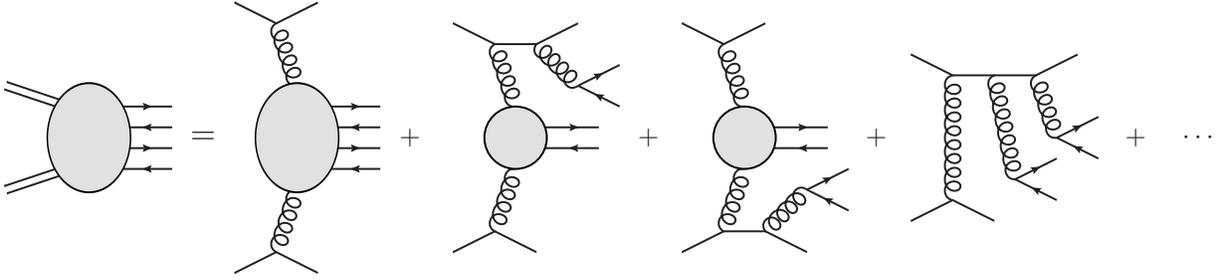,width=\linewidth}
\caption{\label{gaugeterms}Some terms in the expansion of the amplitude in terms of the eikonal propagators. The eikonal quarks are denoted by lines without arrows. The double lines on the left hand side represent the off-shell gluons. This expansion does not represent the organization of the calculation, and only gives an impression of which graphs are included.}
\end{center}
\end{figure}
As mentioned in the introduction, the scattering amplitudes with off-shell initial state gluons are constructed using the formalism of Ref.~\cite{vanHameren:2012if}, in which off-shell gluons are represented by eikonal quark-antiquark pairs in order to arrive at gauge invariant amplitudes. Figure~\ref{gaugeterms} gives an idea of what kind of graphs are included.
The Feynman rules related to the eikonal quark-antiquark-gluon vertex and eikonal propagator are
\begin{equation}
\raisebox{-3.5ex}{\epsfig{figure=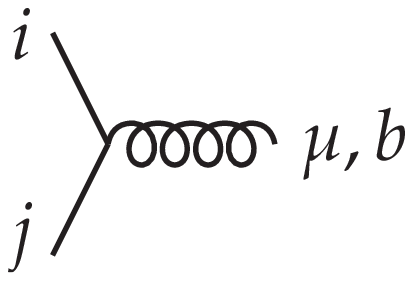,width=14.8ex}}
  \;=\; -\mathrm{i}p_A^\mu T^b_{i,j}
\quad\quad,\quad\quad
\raisebox{-1.5ex}{\epsfig{figure=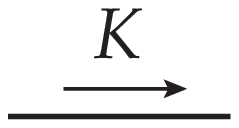,width=8.5ex}}
  \;=\; \frac{\mathrm{i}}{p_A\!\cdot\!K}
\quad\quad,
\end{equation}
where $p_A$ is the longitudinal momentum associated with the off-shell gluon.
The external eikonal quark-antiquark pairs carry fundamental color indices, say $i,j$.
It was noted in Ref.~\cite{Bury:2015dla} that the amplitude is traceless with respect to these indices, so an adjoint color index can be assigned to the off-shell gluon by contracting the amplitude with $\sqrt{2}T^a_{ij}$.
The squared amplitude summed over colors gives the same result. Denoting by $\mathcal{M}^{a}$ the amplitude with the color of one off-shell gluon highlited explicitly
we have
\begin{equation}
\sum_{a}\left|\mathcal{M}^{a}\right|^2
=
\sum_{a}\left|\sqrt{2}\sum_{i,j}\mathcal{M}_{ij}T^a_{ij}\right|^2
=
\sum_{i,j,k,l}\mathcal{M}_{ij}\mathcal{M}_{kl}^*\left(\delta_{ik}\delta_{lj}-\frac{1}{N_{c}}\delta_{ij}\delta_{kl}\right)
=\sum_{i,j}\left|\mathcal{M}_{ij}\right|^2
~.
\end{equation}
The first term on the right hand side of Fig.~\ref{gaugeterms} contains the "actual off-shell gluons" as virtual gluons, representing complete propagators.
This term would diverge if $k_{1t}^2\to0$ and/or $k_{2t}^2\to0$, so the whole amplitude has to be multiplied with $\sqrt{k_{1t}^2k_{2t}^2}$
to reproduce correct collinear limit.

\begin{figure}
\begin{center}
\epsfig{figure=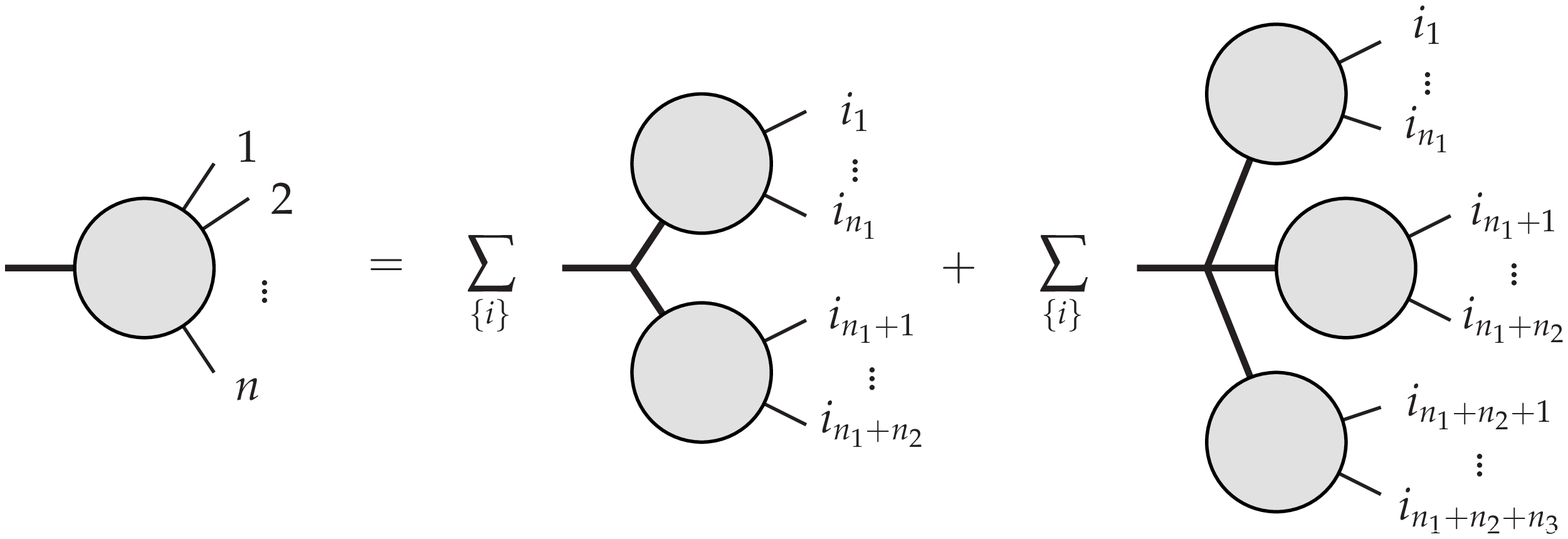,width=0.8\linewidth}
\caption{\label{DysonSchwinger}Dyson-Schwinger recursion for off-shell currents. The thick lines represent off-shell (virtual) particles and the thin lines represent on-shell external particles. The sum is over all partitions of these external particles over the different blobs  and all flavors for the virtual particles that are allowed according to the Feynman rules.}
\end{center}
\end{figure}

\begin{figure}
\begin{center}
\epsfig{figure=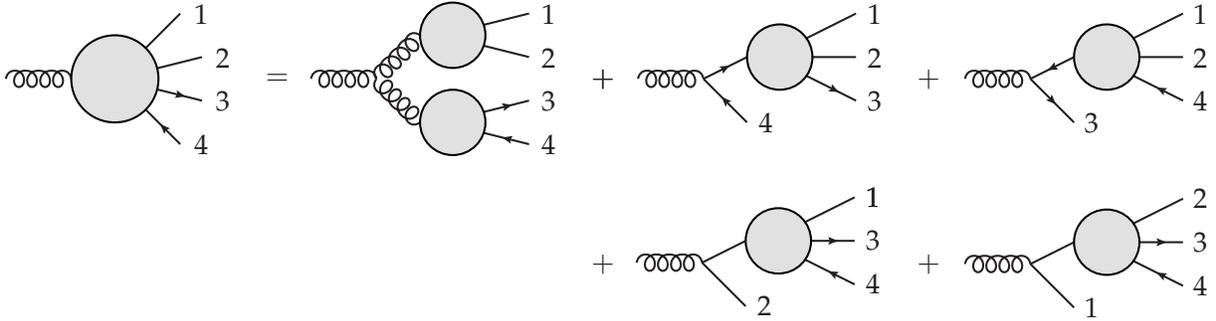,width=\linewidth}
\caption{\label{DSexample}An explicit example of one Dyson-Schwinger recursive step for a certain off-shell current.}
\end{center}
\end{figure}

The calculation has been performed with the help of A Very Handy LIBrary~\cite{Bury:2015dla}.
In this Fortran library, scattering amplitudes are calculated numerically as a function of the external four-momenta via Dyson-Schwinger recursion~\cite{Caravaglios:1995cd}.
It is a recursion of off-shell currents, which automatically factorizes the calculation of the sum of all Feynman graphs such that the multiplications represented by vertices are executed only once for each vertex, while such vertex may occur in several graphs, for identical kinematics.
This recursion is sketched in Fig.~\ref{DysonSchwinger} and Fig.~\ref{DSexample}. The auxiliary eikonal quarks and anti-quarks are treated as external particles, so eventually an eight-point amplitudes are calculated.
AVHLIB allows for various choices of the representation of the external helicities and colors.
These include both the color-ordered representation~\cite{Kanaki:2000ms,Maltoni:2002mq}, with exact summation over color, and the color-dressed representation~\cite{Papadopoulos:2005ky,Duhr:2006iq}, with Monte Carlo summation for large multiplicities.
Helicity configurations can be summed exactly, or, again for large multiplicities, treated in a Monte Carlo approach, both discrete and with continuous random polarizations~\cite{Draggiotis:1998gr}.
The library includes a full Monte Carlo program with an adaptive phase space generator~\cite{vanHameren:2007pt,vanHameren:2010gg} that deals with the integration variables related to both the initial-state momenta and the final-state momenta.

The program can also conveniently generate a file of unweighted events, which approach was used for the analysis presented in this paper.
In the present calculation we use: $\mu_f^2 = (\sum_{i}^4 m_{i,t})^2$ as the factorization scale
and $m_c$ = 1.5 GeV in both $k_t$-factorization and in the reference
collinear-factorization calculations.
Uncertainties related to the choice of the parameters were discussed
e.g. in Ref.~\cite{vanHameren:2014ava} and will be not considered here.
Here we wish to concentrate on the relative effect and modifications
with respect to the results of the collinear-factorization calculations presented already in 
the literature \cite{Schafer:2012tf,vanHameren:2014ava}.

\section{First Results}

In this section we wish to compare the new results of the $k_t$-factorization approach to
those obtained by us in Ref.~\cite{vanHameren:2014ava} in the collinear-factorization approach.

\begin{figure}[!h]
\begin{minipage}{0.47\textwidth}
 \centerline{\includegraphics[width=1.0\textwidth]{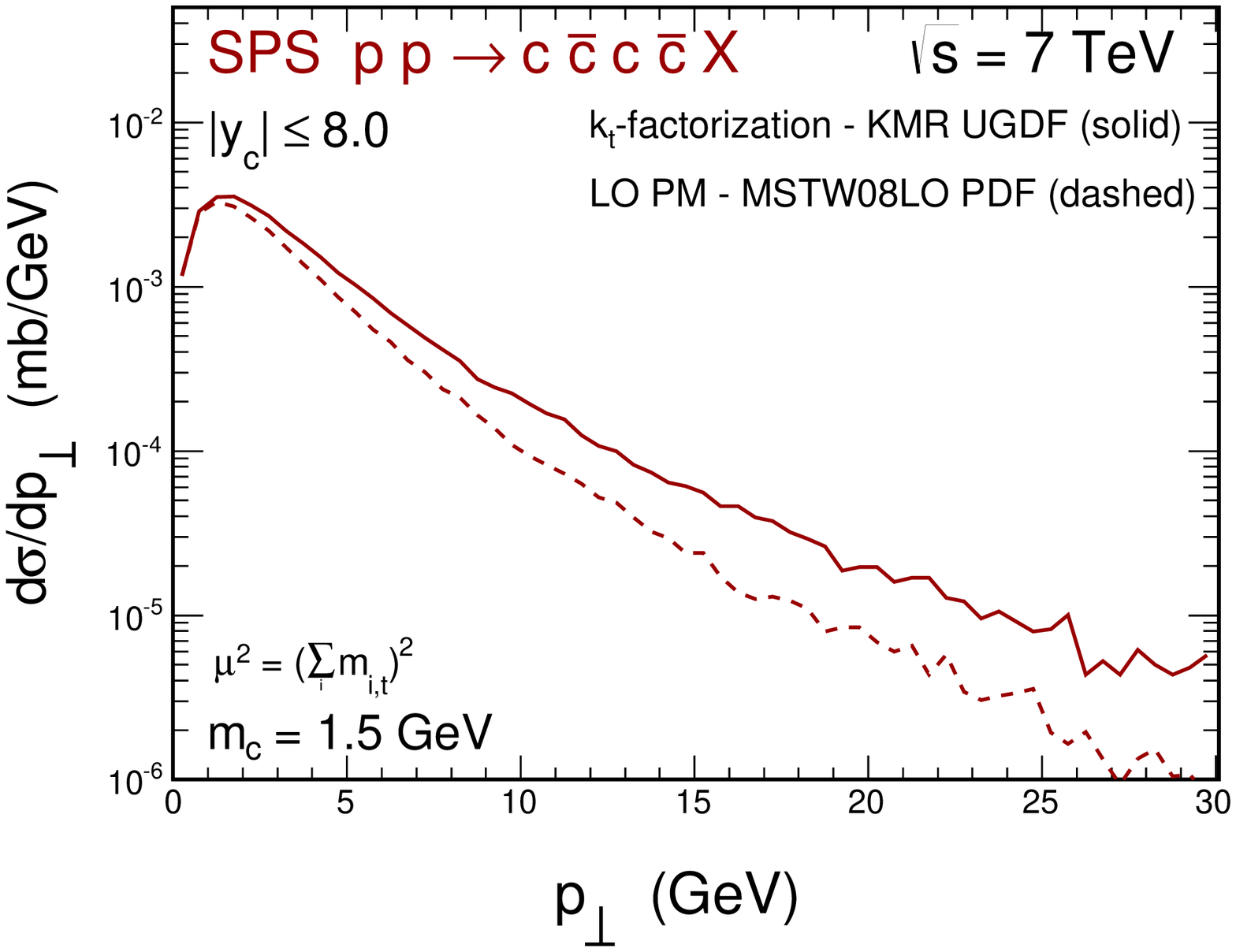}}
\end{minipage}
\hspace{0.5cm}
\begin{minipage}{0.47\textwidth}
 \centerline{\includegraphics[width=1.0\textwidth]{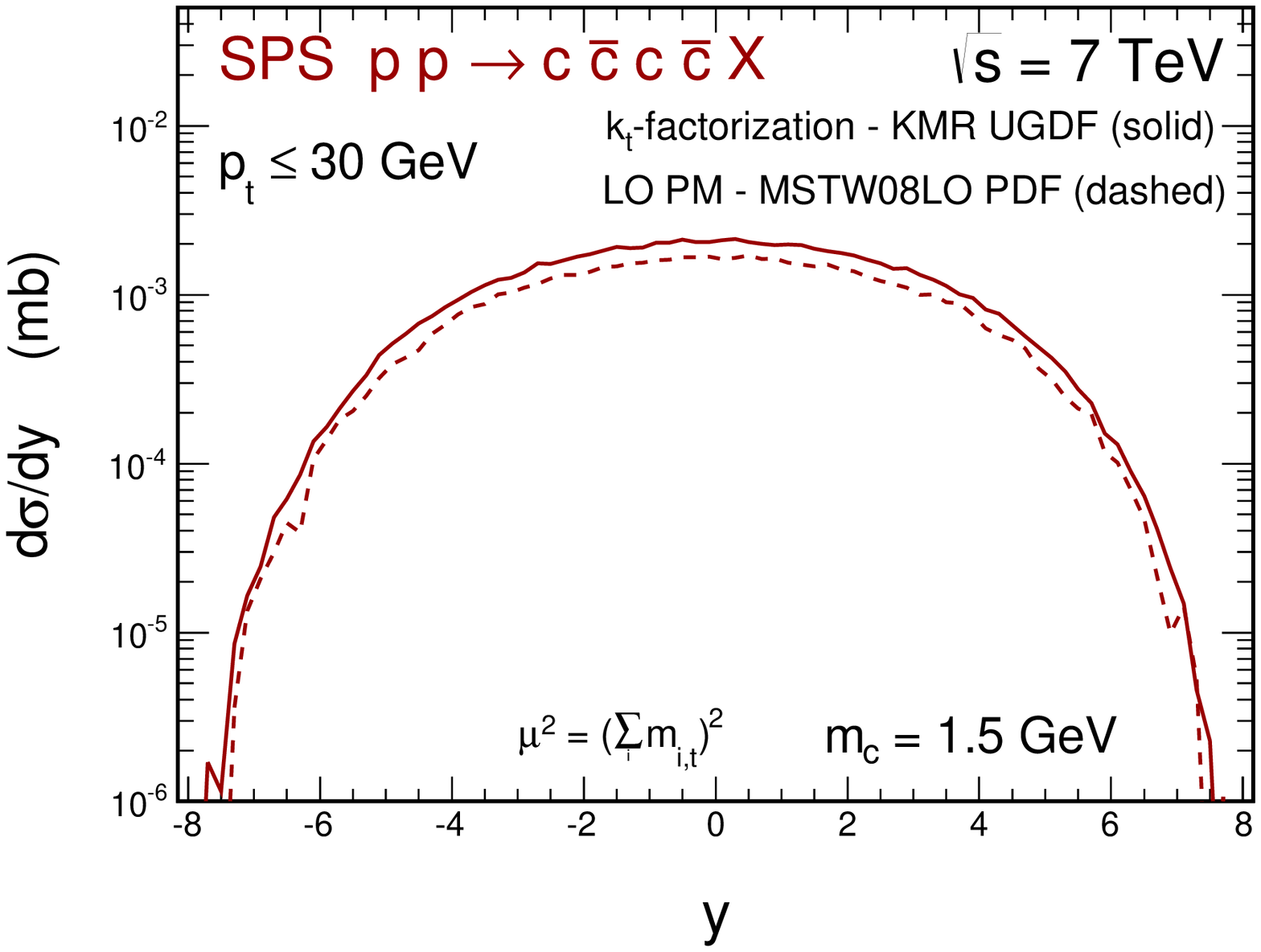}}
\end{minipage}
\caption{
\small Distributions in $c$ quark ($\bar c$ antiquark) transverse momentum
(left panel) and rapidity (right panel). The $k_t$-factorization result
(solid line) is compared with the collinear-factorization result (dashed
line).
}
 \label{fig:dsig_dpt_dy}
\end{figure}

In Fig.~\ref{fig:dsig_dpt_dy} we show standard single
particle distributions in charm quark/antiquark transverse momentum
(left panel) and its rapidity (right panel). We predict an enhancement
of the cross section at large transverse momenta of $c$ or $\bar{c}$ compared to 
the collinear-factorization approach.
The rapidity distributions in both approaches are rather similar
(see the left panel of the figure).

\begin{figure}[!h]
\begin{minipage}{0.47\textwidth}
 \centerline{\includegraphics[width=1.0\textwidth]{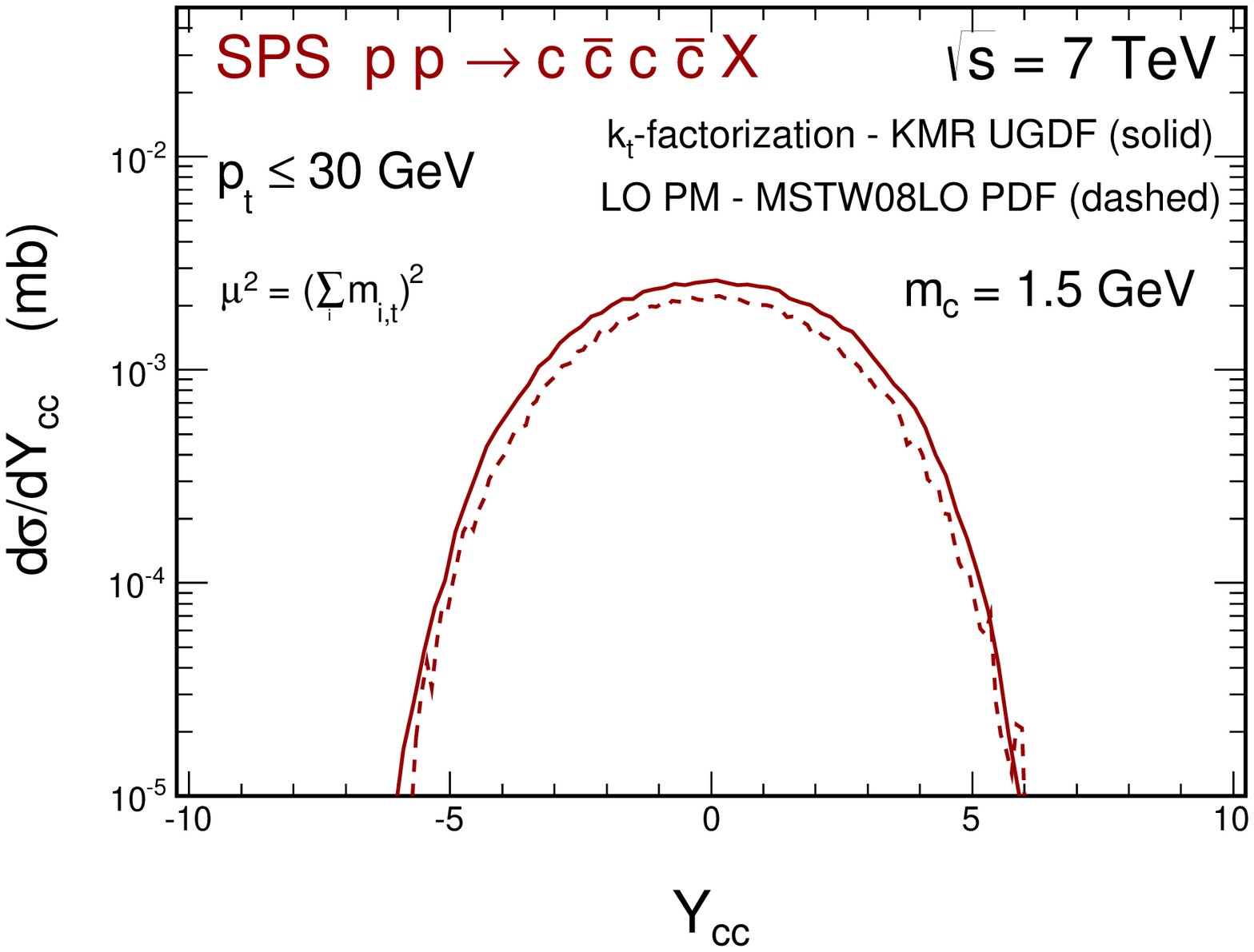}}
\end{minipage}
\hspace{0.5cm}
\begin{minipage}{0.47\textwidth}
 \centerline{\includegraphics[width=1.0\textwidth]{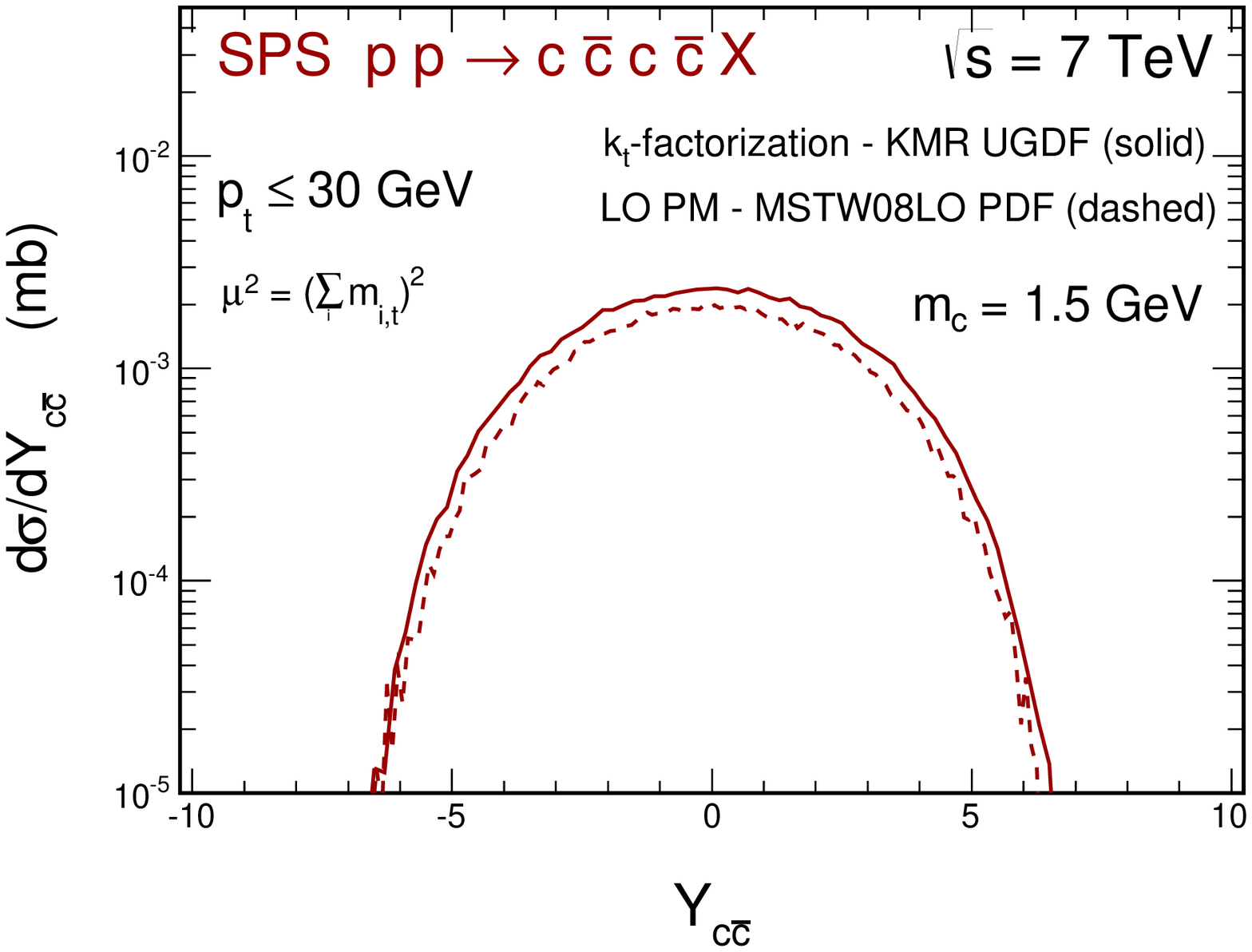}}
\end{minipage}
\caption{
\small Distributions in rapidity $Y_{cc} = (y_c + y_c)/2$ (left panel) and $Y_{c \bar c} = (y_c + y_{\bar c})/2$
(right panel). The meaning of the curves is the same as 
in Fig.~\ref{fig:dsig_dpt_dy}.
}
 \label{fig:dsig_dysum}
\end{figure}

Distributions in rapidity of the $c c$ (or $\bar c \bar c$)
and $c \bar c$, defined as $Y_{cc} = (y_c + y_c)/2$ and $Y_{c \bar c} = (y_c + y_{\bar c})/2$ respectively, are shown 
in Fig.~\ref{fig:dsig_dysum}. The distributions are much narrower than 
those for single quark/antiquark which reflects the fact
that the two different $c$ quarks (or two different $\bar c$ antiquarks) 
have typically different rapidities. The discussed distributions in $y_c$ and $Y_{cc}$
would be identical only if $y_{c,1} = y_{c,2}$ (strong rapidity correlations).
This will become clearer when inspecting rapidity difference in the next plots.

\begin{figure}[!h]
\begin{minipage}{0.47\textwidth}
 \centerline{\includegraphics[width=1.0\textwidth]{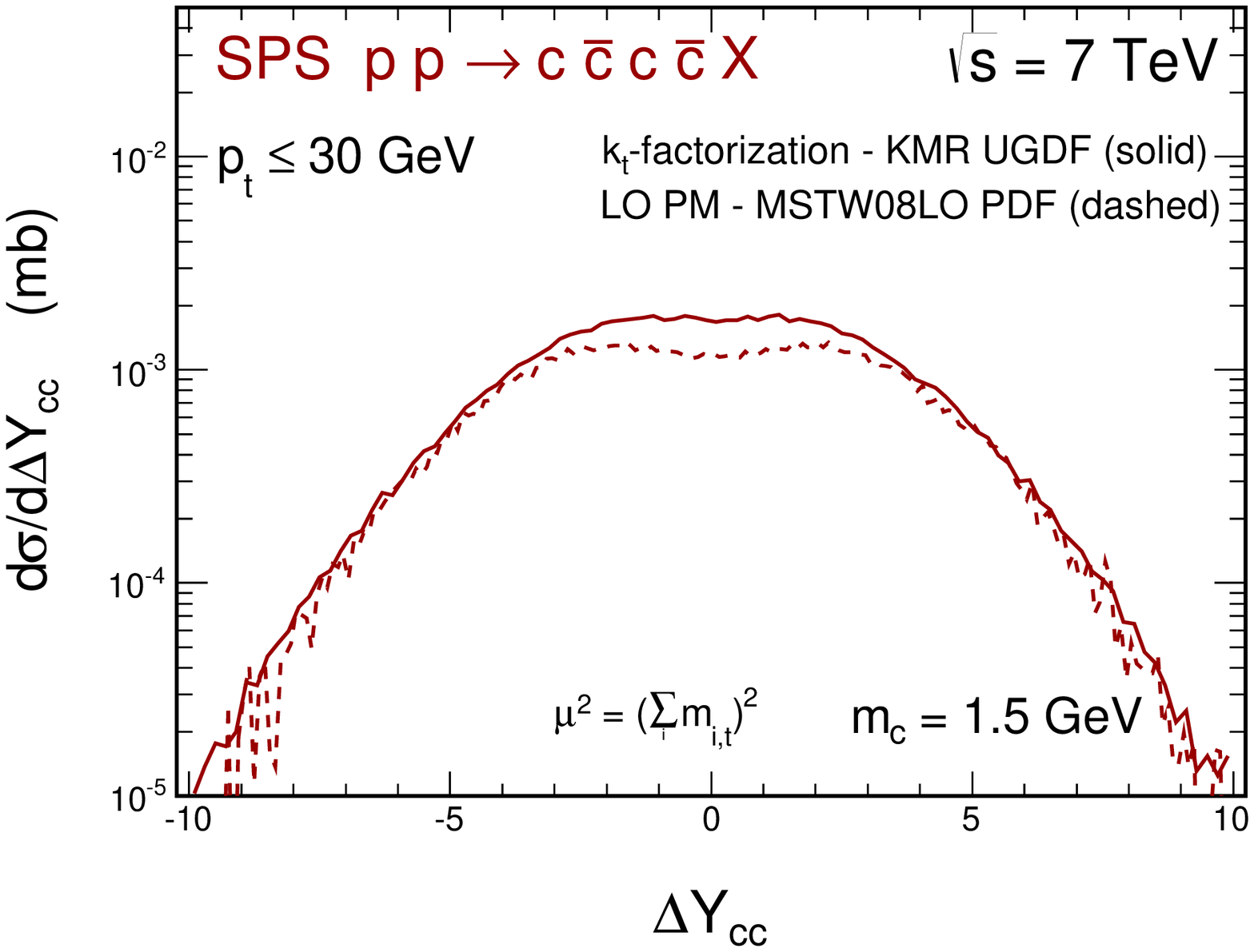}}
\end{minipage}
\hspace{0.5cm}
\begin{minipage}{0.47\textwidth}
 \centerline{\includegraphics[width=1.0\textwidth]{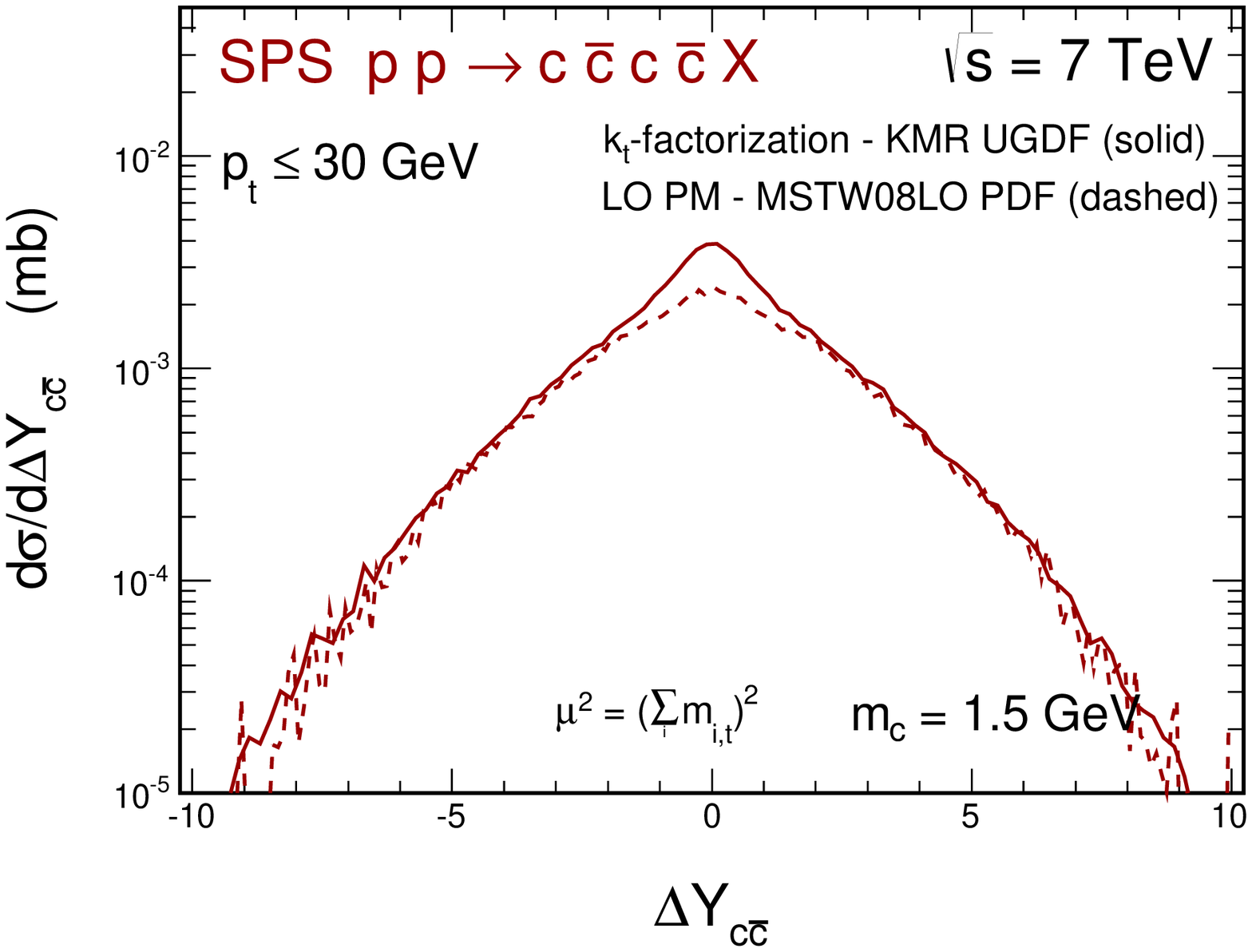}}
\end{minipage}
\caption{
\small Distributions in the difference of rapidities 
$\Delta Y_{cc} = y_c - y_c$ (left panel) and $\Delta Y_{c \bar c} = y_c - y_{\bar c}$
(right panel). The meaning of the curves is the same as 
in Fig.~\ref{fig:dsig_dpt_dy}.
}
 \label{fig:dsig_dydiff}
\end{figure}

Similar distributions but for rapidity distance between two $c$ quarks 
(or two $\bar c$ antiquarks) and between  $c$ and $\bar c$
are shown in Fig.~\ref{fig:dsig_dydiff}. On average the distance
between $c$ and $c$ is larger than that for $c$ and $\bar c$.
This can be understood easily in the high-energy approximation 
discussed in Ref.~\cite{Schafer:2012tf} by inspecting the contributing diagrams.
Some enhancement at small rapidity separations can be observed in the $k_t$-factorization approach
compared to the collinear approach.

\begin{figure}[!h]
\begin{minipage}{0.47\textwidth}
 \centerline{\includegraphics[width=1.0\textwidth]{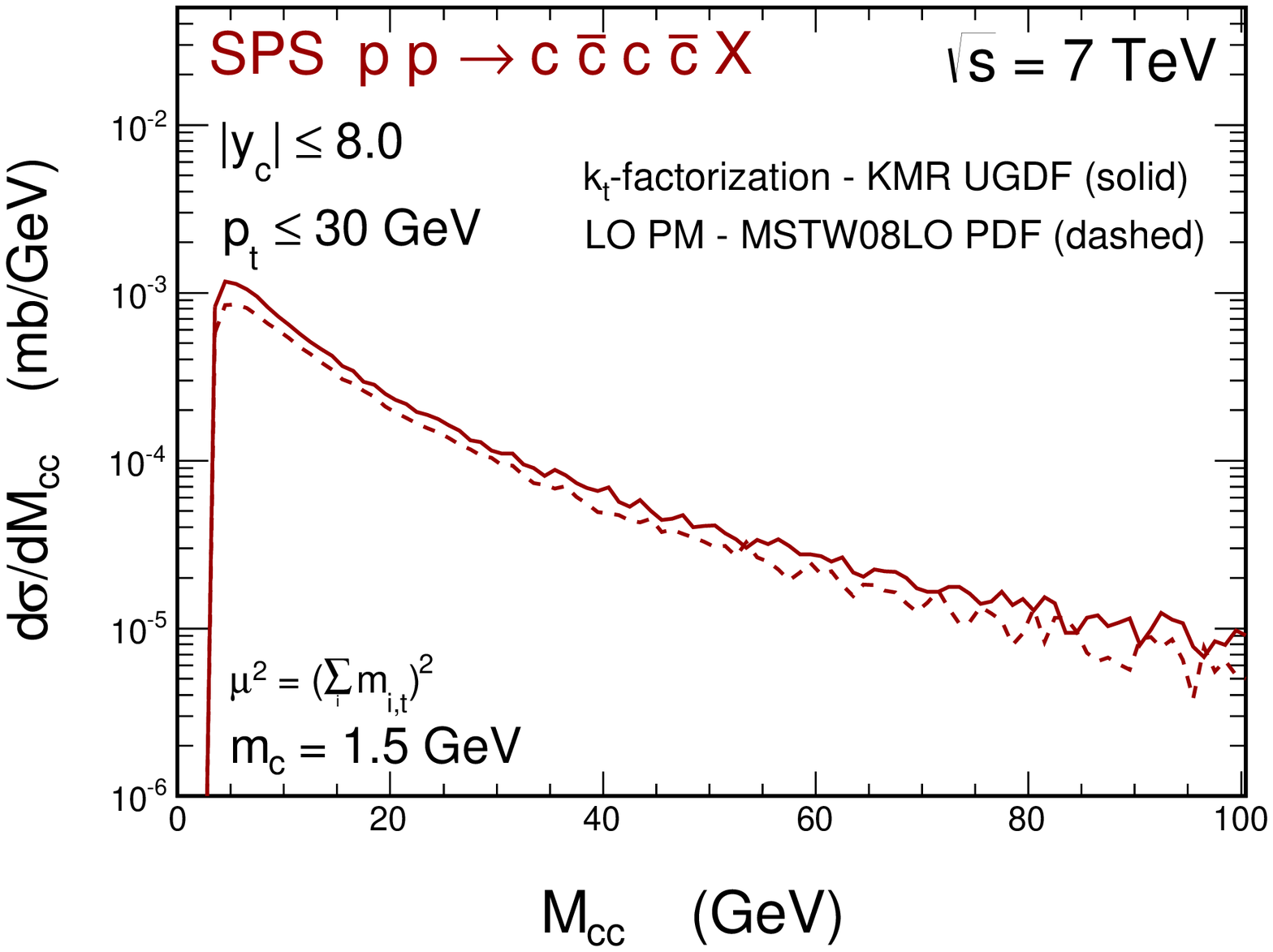}}
\end{minipage}
\hspace{0.5cm}
\begin{minipage}{0.47\textwidth}
 \centerline{\includegraphics[width=1.0\textwidth]{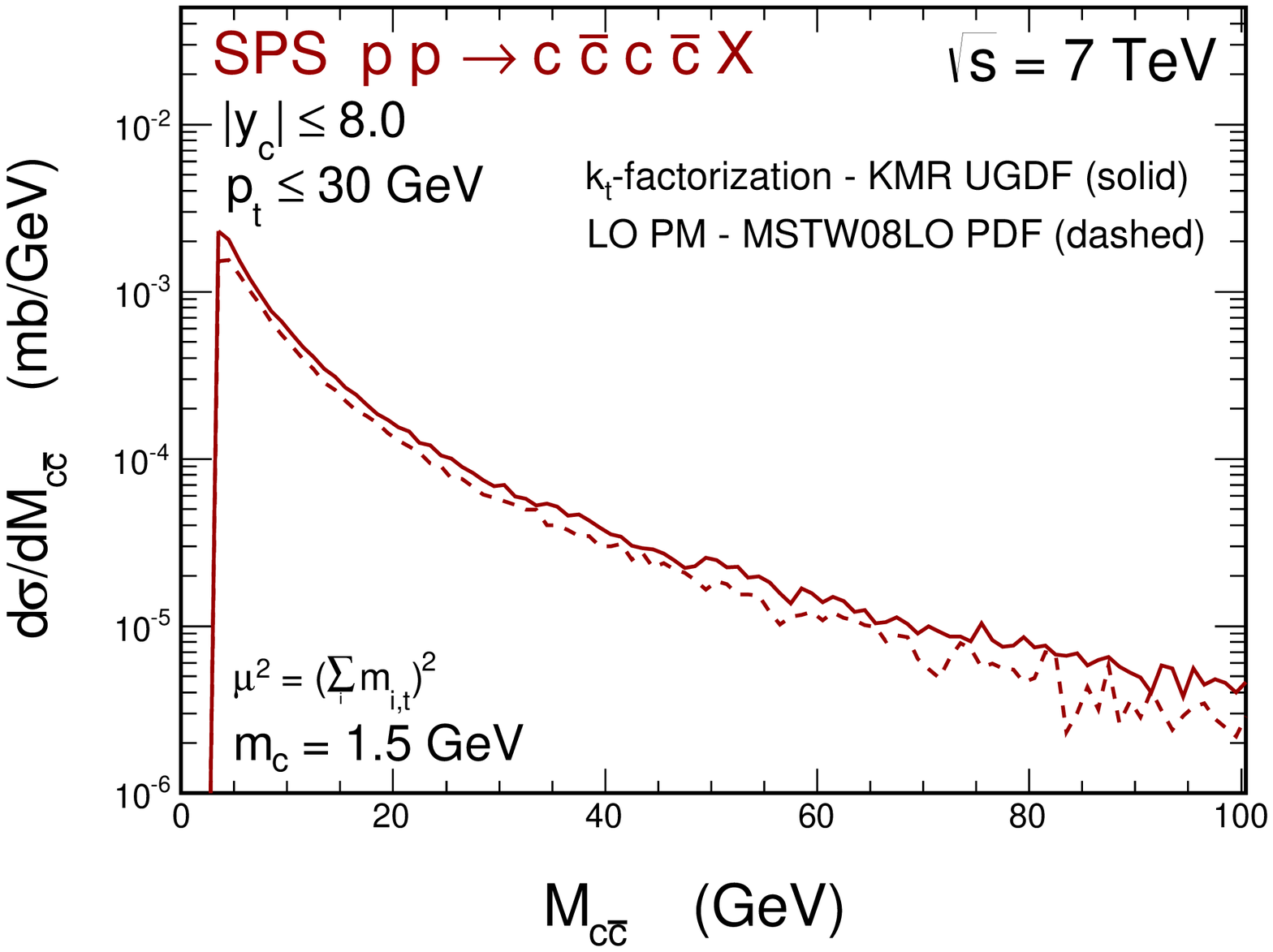}}
\end{minipage}
\caption{
\small Invariant mass distrtributions in $M_{cc}$ (left panel)
and $M_{c \bar c}$ (right panel).
The meaning of the curves is the same as in Fig.~\ref{fig:dsig_dpt_dy}.
}
 \label{fig:dsig_dMcc}
\end{figure}

The distributions in rapidity distance are strongly correlated
with $M_{cc}$ or $M_{c \bar c}$ distributions shown in Fig.~\ref{fig:dsig_dMcc}.
Those distribution are, however, difficult to measure as
rather mesons are measured and not quarks or antiquarks.

\begin{figure}[!h]
\begin{minipage}{0.47\textwidth}
 \centerline{\includegraphics[width=1.0\textwidth]{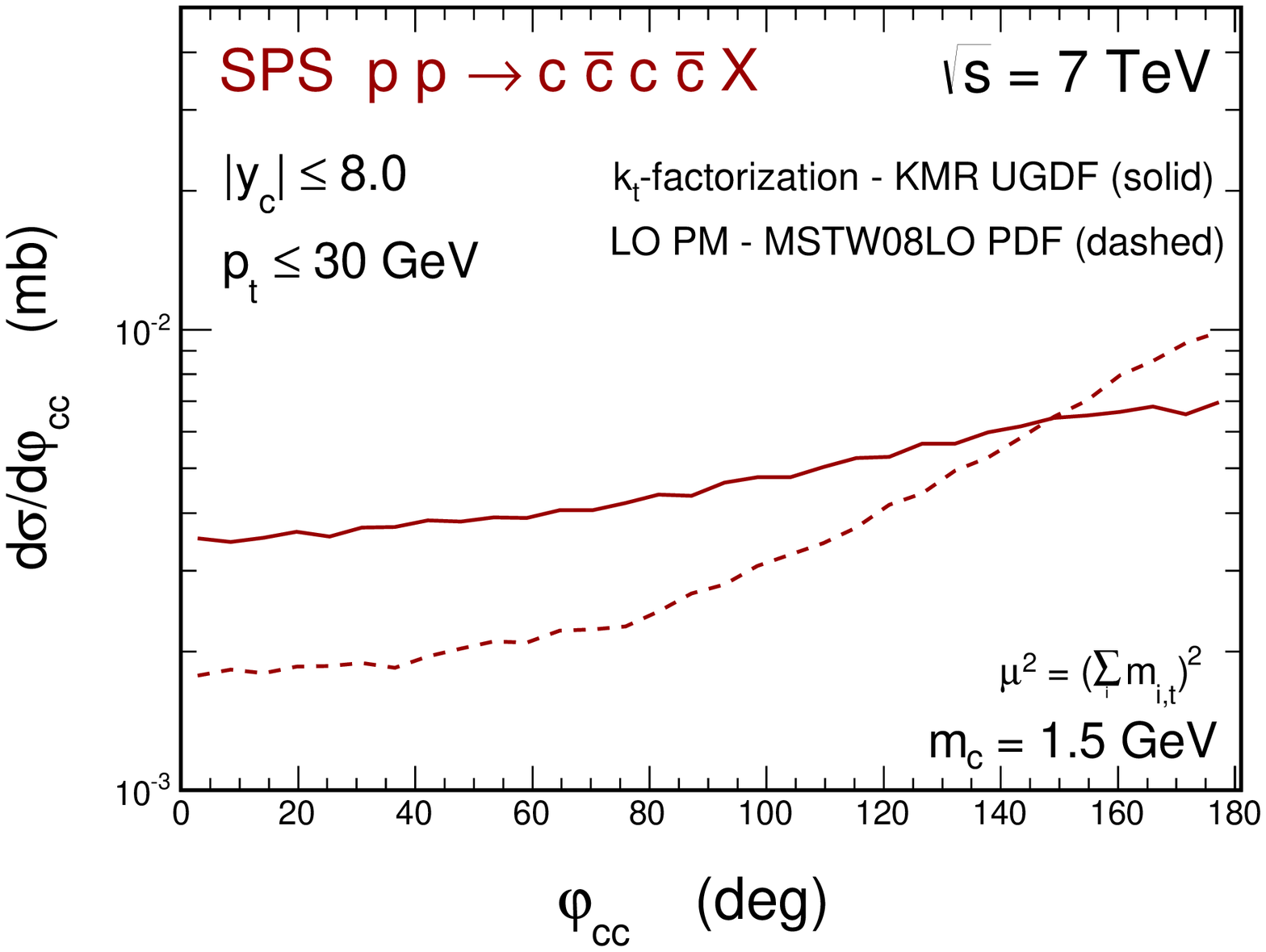}}
\end{minipage}
\hspace{0.5cm}
\begin{minipage}{0.47\textwidth}
 \centerline{\includegraphics[width=1.0\textwidth]{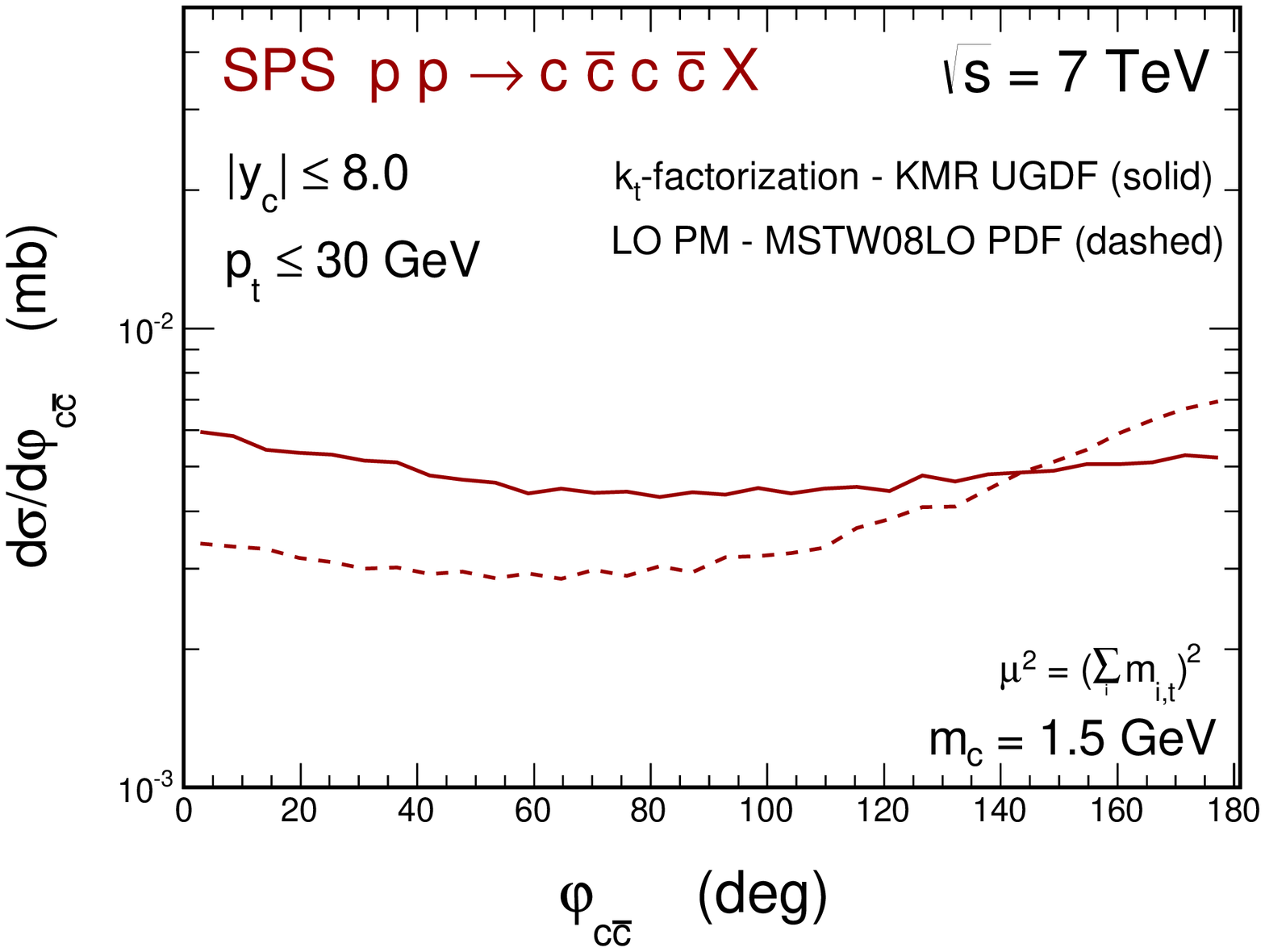}}
\end{minipage}
\caption{
\small Azimuthal angle correlations between two $c$ quarks (left panel)
and between $c$ and $\bar c$ (right panel).
The meaning of the curves is the same as in Fig.~\ref{fig:dsig_dpt_dy}.
}
 \label{fig:dsig_dphi}
\end{figure}

Quite interesting are azimuthal angle correlations between
$c$ and $c$ or $c$ and $\bar c$. The corresponding distributions
are shown in Fig.~\ref{fig:dsig_dphi}.
We note much bigger decorrelation of two $c$ quarks or $c$ and $\bar c$ 
in the $k_t$-factorization approach compared to the collinear approach.
This is due to explict account of gluon virtualities (transverse momenta).
We will return to this point when discussing azimuthal correlations between mesons at the end of this section.
 
\begin{figure}
\includegraphics[width=8cm]{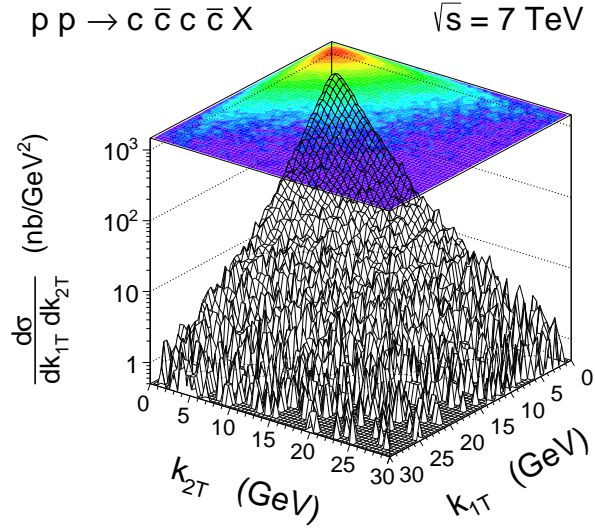}
\caption{
\small Two-dimensional distribution in transverse momenta
of initial gluons in the $p p \to c \bar c c \bar c$ SPS process
at $\sqrt{s}$ = 7 TeV.
}
\label{fig:dsig_dk1tdk2t}
\end{figure}

Next we wish to visualize the regions of the transverse momenta
of initial gluons that give sizeable contribution to 
the SPS $p p \to c \bar c c \bar c X$ cross section. 
In Fig.~\ref{fig:dsig_dk1tdk2t} we show a two-dimensional
distribution in intial-gluon transverse momenta.
The dependence on $k_{1t}$ and $k_{2t}$ shown in the figure is determineated
by the UGDF used in the calculation as well as by the dependence of the matrix element on
$k_{1t}$ and $k_{2t}$. Other models of unintegrated gluon distributions would give
different dependencies.
Clearly we get large contributions from the regions far from
the collinear case ($k_{1t}$ = 0 and $k_{2t}$ = 0). 
This has of course consequences for other observables discussed above
through the dependence of the matrix element on the gluon transverse
momenta  $\overline{|{\cal M}(k_{1t},k_{2t})|^2}$ and its correlation
with other kinematical variables.

We will not discuss in the present letter the correlations between
the gluon virtualities (or their transverse momenta) and other kinematical
variables related to the charm quarks and antiquarks.

\begin{figure}[!h]
\begin{minipage}{0.47\textwidth}
 \centerline{\includegraphics[width=1.0\textwidth]{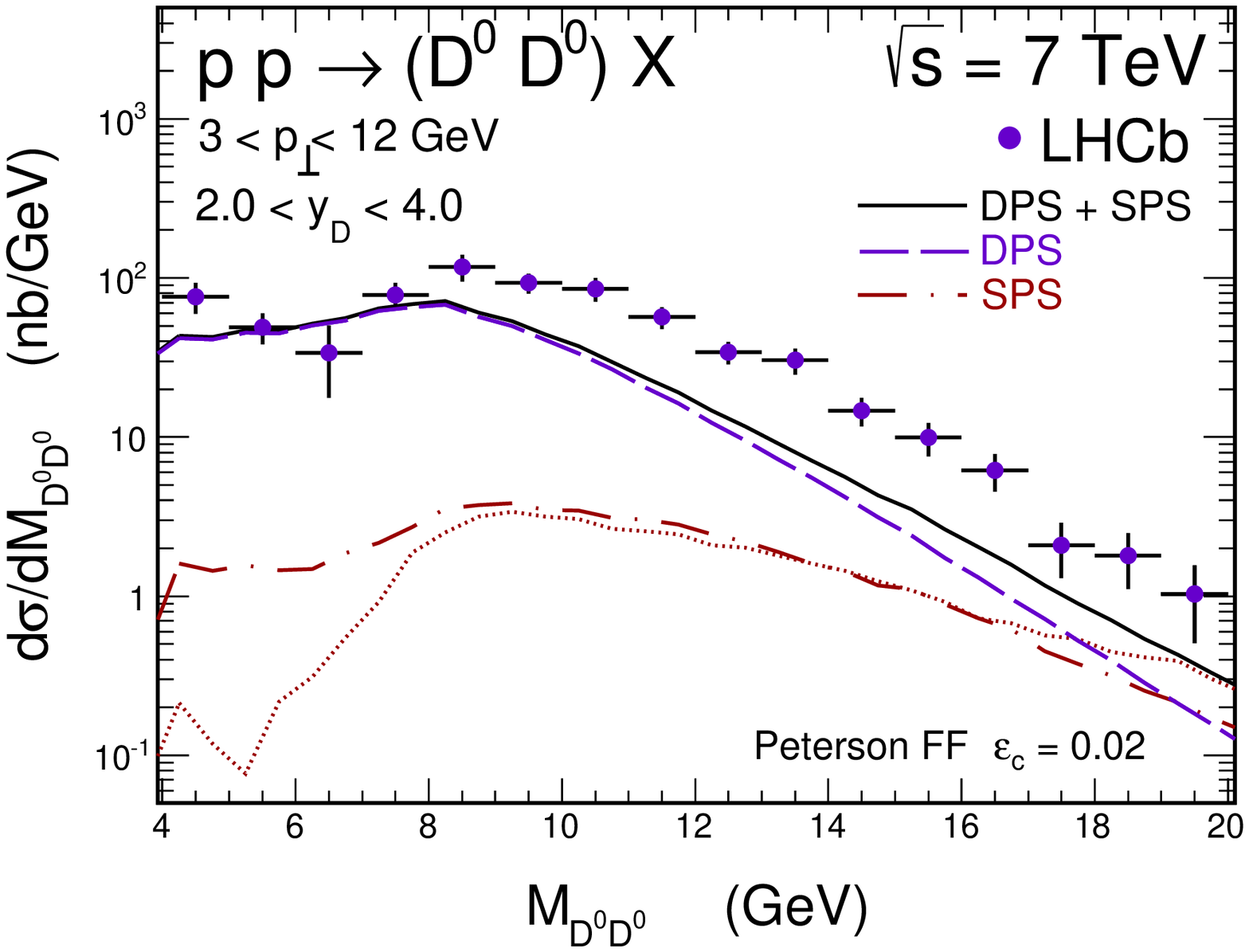}}
\end{minipage}
\hspace{0.5cm}
\begin{minipage}{0.47\textwidth}
 \centerline{\includegraphics[width=1.0\textwidth]{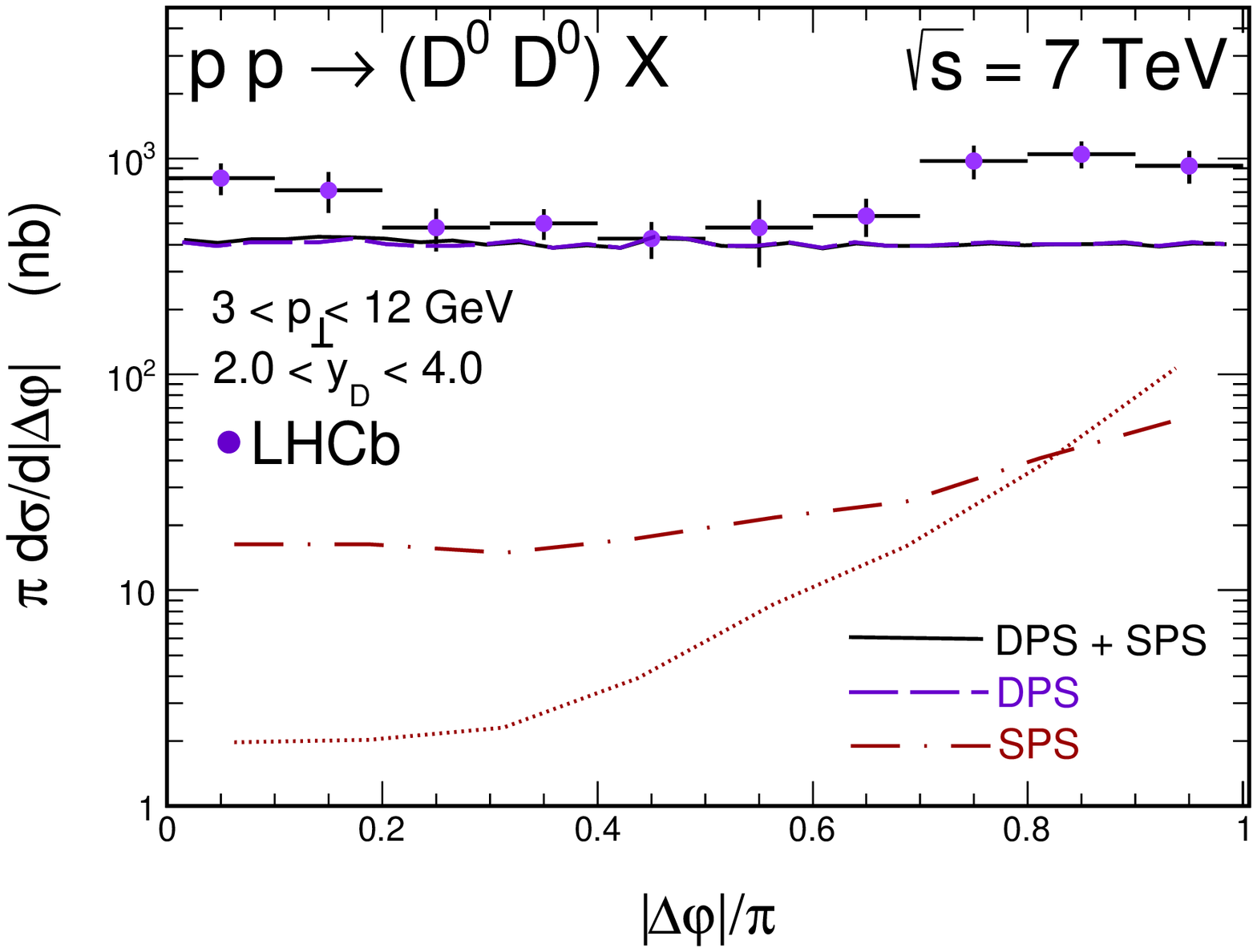}}
\end{minipage}
\caption{
\small Distributions in $D^0D^0$ invariant mass (left) and in azimuthal angle between both $D^0$'s (right) within
the LHCb acceptance. The DPS contribution (dashed line) is compared with the SPS one calculated within the $k_t$-factorization approach (dashed-dotted line).
The SPS result from our previous studies \cite{vanHameren:2014ava}, calculated in the LO collinear-factorization approach, is also shown here (dotted line). 
}
 \label{fig:LHCb_D0D0}
\end{figure}

So far we have considered production of $c \bar c c \bar c$ quarks/antiquarks.
As discussed in our previous paper \cite{vanHameren:2014ava} such a final states may lead
to the production of two $D$ mesons, both containing $c$ quarks or both containing
$\bar c$ antiquarks which is not possible e.g. for the $c \bar c$ final-state case.
As explained in Ref.~\cite{vanHameren:2014ava} the DPS gives cross sections
very similar to those measured by the LHCb collaboration \cite{Aaij:2012dz}.
How important is the SPS contribution discussed in this paper, calculated here in the $k_t$-factorization,
is shown in Fig.~\ref{fig:LHCb_D0D0}. For comparison we show also SPS results
calculated in collinear-factorization approach \cite{vanHameren:2014ava}. The two approaches
give somewhat different shapes of correlation observables, inspite that the integrated
cross sections are rather similar as discussed already at the parton level.
Our results, so far the most advanced in the literature as for as the SPS contribution is considered,
are not able to explain discrepancy between DPS contribution and the LHCb experimental
data. If the discrepancies are due to simplifications in the treatment of DPS
requires further studies including for example spin and flavour correlations.
Some works in this direction already started \cite{Mulders2015}.

\section{Conclusions}

In the present paper we have made a first calculation of the cross section
for $p p \to c \bar c c \bar c X$ in the $k_t$-factorization approach,
i.e.\ focussing on single parton scattering process.
This is a first $2 \to 4$ process for which
$k_t$-factorization is applied. In this calculation we have
used the Kimber-Martin-Ryskin unintegrated gluon distribution(s)
which effectively takes into account the dominant higher-order corrections.
The off-shell matrix element was calculated using a new technique
developed recently in Krak\'ow.

The results of the $k_t$-factorization approach were compared
with the results of the collinear-factorization approach.
In general, the $k_t$-factorization results are only slightly bigger than
those for collinear approach.
An exception is the transverse momentum distribution above 10 GeV where
a sizeable enhancement has been observed.
Inclusion of gluon virtualities leads to a decorrelation in azimuthal
angle between $c$ and $c$ or $c$ and $\bar c$.

Since the cross section is in general very similar as for
the collinear-factorization approach we conclude that the
$c \bar c c \bar c$ final state at the LHC energies is dominantly 
produced by the double parton scattering as discussed in our recent
papers,
and the SPS contribution, although interesting by itself, is rather small.
A comparison to predictions of double-parton scattering results
and recent LHCb data for azimuthal angle correlations between $D^0$ and $D^0$
or $\bar{D}^0$ and ${\bar D}^0$ mesons strongly suggests that the assumption 
of two fully independent DPS 
($g g \to c \bar c \otimes g g \to c \bar c$) may be too approximate
or even not valid. Some possible reasons were discussed in
Ref.~\cite{Mulders2015}. The effect found there is, however, too small to
explain a rather large effect observed by the LHCb collaboration.
This remains a challenge for future theoretical studies and should be confirmed
by the LHCb collaboration at $\sqrt{s} = 13$, $14$ TeV.

\vspace{1cm}

{\bf Acknowledgments}

This study was partially supported by the Centre for
Innovation and Transfer of Natural Sciences and Engineering Knowledge in
Rzesz{\'o}w. Graphs were drawn with the help of Jaxodraw.


\end{document}